\documentstyle{sup}
\input psfig

\newcommand{\be}{\begin{eqnarray}}
\newcommand{\ee}{\end{eqnarray}}

\begin{document}
\title[{\it Superlattices and Microstructures, Vol. ??, No. ?, 1999}]
{Resonant tunneling through Andreev levels}
\author[{\it Superlattices and Microstructures, Vol. ??, No. ?, 1999}]
{A. Kadigrobov and L. Y. Gorelik
\cr
{\normalsize\it Department of Applied Physics}\cr
{\normalsize\it Chalmers University of Technology and G{\"o}teborg University,
SE-412 96 G{\"o}teborg, Sweden}
\cr
{\normalsize\it and B. Verkin Institute for Low Temperature Physics and 
Engineering, 310164 Kharkov,  Ukraine}
\vspace{10pt}\cr
R. I. Shekhter and M. Jonson\cr
{\normalsize\it Department of Applied Physics}\cr
{\normalsize\it Chalmers University of Technology and G{\"o}teborg University,
SE-412 96 G{\"o}teborg, Sweden}
}

\maketitle
\vspace*{1cm}
\begin{abstract}
We use a semiclassical approach for analysing the tunneling transport through 
a normal conductor in contact with superconducting mirrors. Our analysis of 
the electron-hole propagation along semiclassical trajectories shows that 
resonant transmission through Andreev levels is possible resulting in an
excess, low-energy quasiparticle contribution to the conductance.
The excess conductance oscillates with the phase difference between the 
superconductors having maxima at odd multiples of $\pi$ for temperatures 
much below the Thouless temperature. 
\end{abstract}
\vspace*{.5cm}

\section{Introduction}
Recent experimental and theoretical work on diffusive charge transport in
mesoscopic N/S samples
[1-39]
%
%
have revealed a strong energy dependence of an excess quasiparticle 
contribution
to the low temperature conductance of normal (N) parts in close proximity to
superconductors (S) (for a review see e.g. \cite{Lambert2}). A characteristic
energy scale is set by the Thouless energy $E_{Th}=\hbar D/L_S^2$ 
below
which a re-entrance to normal conduction is seen as the bias voltage or
temperature is lowered
($D$ is the diffusion coefficient, $L_S$ is the distance between the 
superconductors). 
In samples of the Andreev interferometer type with two
N/S interfaces the excess conductance oscillates as a function of the phase
difference $\phi $ between  two superconductors. Conductance maxima occur at
even multiples of $\pi $, their magnitude peaks at $E_{Th}$ and becomes
vanishingly small at low energies. These oscillations have been explained by
Nazarov and Stoof as a ``thermal effect'' \cite{NazarovStoof,VolkovLambert}
and can be  understood in terms of competing contributions from the condensate
wave functions to the density of states or in terms of quasi-particle
trajectories between the N/S interfaces

In Ref.~\cite{GO}, a strong interference
effect due to resonant transmission of quasiparticles through Andreev 
levels was
shown to take place in superconductor-normal metal-superconductor (S/N/S)
heterostructures at temperatures corresponding to energies much
below the ballistic Thouless energy $E_{Th}=\hbar v_F/L_S$ ($v_F$ is the 
Fermi velocity).
The effect results in giant peaks in the conductance ---
proportional to the number $N_{\perp}$ of conducting transverse modes ---
whenever the phase difference $\phi $ between
the superconductors is an odd multiple of $\pi$. 
Arguments were presented in Ref.~\cite{GOdif} for this result to be
valid also in the diffusive diffusive transport regime.

Andreev levels are bound states formed in a normal sample element by
successive Andreev reflections of quasiparticle excitations at two
S/N-interfaces. A peculiar feature of these bound states is that they carry
the supercurrent (if any) between the two superconductors. These levels can 
also carry a normal transport current if the sample is coupled to reservoirs 
of normal electrons \cite{GOdif}. Such a transmission is of a resonant 
type if  the coupling is weak enough not to destroy the
Andreev levels themselves.
This is the case when the reservoirs are coupled through tunneling barriers 
of low transparency. It is important that, in addition, such barriers 
serve as quantum scatterers extended in two dimensions; they split 
quasiclassical electron trajectories incident from the reservoir, or returning
towards the reservoir after having been Andreev reflected from
an N/S boundary. This enables the trajectory of a quasi-particle  with  
 zero excitation energy (measured from the Fermi level) 
%
%
to connect {\em both} S/N-interfaces (necessary for picking up 
information about
$\phi =\phi _1-\phi _2$, $\phi _1$ and $\phi _2$ being the phases of
superconductors 1 and 2 ) and a reservoir (to affect the current) as shown in
Fig.~1.

In this work, which is based on an analysis of the quasiparticle 
trajectories in 
a disordered  normal conductor weakly coupled to reservoirs,  we calculate 
how the two-terminal 
conductance between two normal resrvoirs depend on both
the superconductor phase difference and temperature. We pay
special attention to non-Andreev (normal) reflections at the N/S boundaries.
Including these, the  character of the resonance is changed shifting the
maximum of the amplitude of conductance oscillations to nonzero
temperatures.  In our concluding section we  discuss recent experimental 
results
of Ref.~\cite{Rais}, where an anomalous low-temperature behavior of the 
conductance was observed.

\begin{figure}
\vspace{3mm}
\centerline{\psfig{figure=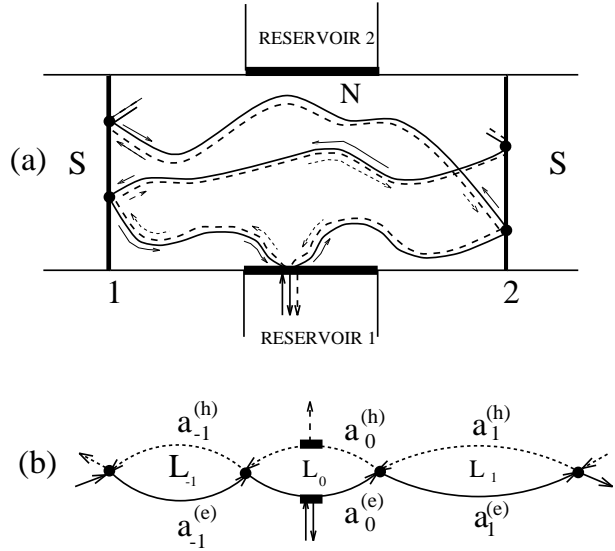,width=8 cm}}
\caption{ \label{fig:one}
(a) Structure analyzed in the text, with superconducting (S) and disordered 
normal (N) elements labeled. The normal element is coupled to two reservoirs
through  tunnel barriers shown as thick black lines. Semiclassical 
trajectories of electron- and hole-like quasiparticle excitations are indicated
by solid  and dashed lines, respectively.  
Sections of the trajectories going between 
N/S boundaries 1 and 2 are connected by Andreev and normal 
reflections at the points shown as black dots. An electron incident 
from the resorvoir (solid up-arrow) may be reflected back either as an 
electron 
(solid down-arrow) or as a hole (dashed arrow).
(b) Sequence of scattering events along the trajectory shown in (a). $L_n$  is
the length of the trajectory  in  section $n$;  $a_n^{(e,h)}$ are the
amplitudes of the corresponding electron and hole excitations at the Fermi
energy ($n=0, \pm 1, \ldots$).  }
\end{figure}

\section{Formulation of the problem}
Figure~1 shows the geometry of the sample under consideration. 
A mesoscopic normal region in contact with two superconductors is coupled to
two reservoirs of normal electrons 
through  tunneling potential barriers of low transparency. 
The quasiparticle motion in the normal region is assumed to be affected by 
a smoothly varying disorder  potential and is treated semiclassically. 

According to the  Landauer-Lambert formula \cite{Lambert1} the conductance of 
the system corresponding to a current between two normal electron
reservoirs can be written in terms of the probability $R_a(E)$ for electrons
impinging from one reservoir with energy $E$ to be reflected as holes back
into the same reservoir and in terms of the probability $T_o(E)$ for them to
be transmitted to  the other reservoir as electrons. For the case of
two equal barriers one has
\begin{equation}
G=\frac{2e^2}{h} \int \left(-\frac{\partial f_0 (E)}{\partial E}\right) 
\left(T_0(E)+R_a(E)\right) dE \, ,
\label{cond}
\end{equation}
where $f_0(E)$ is the Fermi distribution function. Our objective is to 
calculate
the  "electron-hole transmission" probability $R_a(E)$ in Eq.~(\ref{cond}). 
This is
sufficient, since it can be shown that due to the destructive interference
among the relevant trajectories, $T_0(E)$ is only weakly affected by the
superconductor phase difference $\phi$. Hence the effective transmission
probability $T(E)$ for our problem, to be calculated below, is 
given by $R_a(E)$. 

An electron impinging on the lower potential barrier in Fig.~1 from the
reservoir gets, after possibly interacting with the barrier and the two S/N
interfaces in the N-region,  backscattered in the electron and hole channels
with probability amplitudes 
$C^{(e)}_{{\bf p}}({\bf r}_{\parallel})$ and 
$C^{(h)}_{{\bf p}}({\bf r}_{\parallel}$) respectively.
The coordinate ${\bf r}_{\parallel}= (0, y,z)$ lies in the plane of the
barrier, the $x$-axis being perpendicular to the barrier plane. 
Accordingly, the
wave function of the electron  $u({\bf r}_{\parallel})$ and hole $v({\bf
r}_{\parallel})$ in the reservoir can be written as 
\begin{eqnarray} 
\nonumber
u({\bf r})&=&{\rm e}^{i{\bf p}_{\parallel}{\bf r}_{\parallel}} \left({\rm
e}^{ip_x^{(e)} x}+ C^{(e)}_{{\bf p}}({\bf r}_{\parallel}) {\rm
e}^{-ip_x^{(e)}x}\right) \\ v({\bf r})&=&
{\rm e}^{i{\bf p}_{\parallel}^{(h)}{\bf
r}_{\parallel}} C^{(h)}_{{\bf p}}({\bf r}_{\parallel}){\rm e}^{ip_x^{(h)} x} \,
;  \quad\quad\quad
p_x^{(e,h)}=\sqrt{p_F^2 -{\bf    p}_{\parallel}^{(h)2}\pm E}
\label{reserv}
\end{eqnarray}
The probability for electron-hole transmission, which we need to calculate, is
connected with the  amplitude $C^{(h)}_{{\bf p}}({\bf r}_{\parallel})$ by the
relation 
\begin{equation}
T(E)=\frac{1}{(2\pi \hbar)^2}\int_{p_{\parallel}^2/2m 
\leq \epsilon_F}d{\bf p_{\parallel}}\int_{S}d
{\bf r_{\parallel}}|C_{{\bf p}}({\bf r}_{\parallel})|^2 \, ,
\label{intcon}
\end{equation}
where $S$ is the area of the tunnel barrier.

The wave function in the normal region N near the potential barrier is
also characterized by electron and hole components,
\begin{eqnarray}
\nonumber
u({\bf r})&=&{\rm e}^{i{\bf p}_{\parallel}{\bf r}_{\parallel}} 
\left[A^{(e)}_0({\bf r}_{\parallel}){\rm e}^{ip_x^{(e)} x}+
B^{(e)}_0({\bf r}_{\parallel}){\rm e}^{-ip_x^{(e)} x}\right] \\
v({\bf r})&=&{\rm e}^{i{\bf p}_{\parallel}{\bf r}_{\parallel}} 
\left[A^{(h)}_0({\bf r}_{\parallel}){\rm e}^{ip_x^{(h)} x}+
B^{(h)}_0({\bf r}_{\parallel}){\rm e}^{-ip_x^{(h)} x}\right]
\label{norm}
\end{eqnarray}
A slowly varying potential in the N region (on the scale of the Fermi  
wavelength $\lambda_F$) is responsible for the semiclassical nature of the 
wavefunction (\ref{norm}) and the slow spatial variation of the factors
$A^{(e,h)}({\bf r}_{\parallel})$  and  $B^{(e,h)}({\bf r}_{\parallel})$.
The matching conditions at the barrier, which can be expressed in terms of a
unitary $2\times 2$ matrix, describe the coupling between 
$C^{(e,h)}({\bf r}_{\parallel})$  and $A^{(e,h)}({\bf r}_{\parallel})$,  
$B^{(e,h)}({\bf r}_{\parallel})$.  We assume
that these  functions smoothly --- in the semiclassical sense ---
go to zero at the perimeter of  the injector reservoir and are equal to zero in
the  plane $x=0$ outside the injector.

Using the language of semiclassical propagation we now construct the
wavefunction in the N-region (see Fig.~1) for the electron impinging on the
lower barrier from the reservoir. We do so by mapping it to the  wave function 
(\ref{reserv}) at the injector barrier. 
Such a mapping is possible if the disorder potential does not cause any 
noticable  
divergence of a tube of trajectories with the transverse size of the order 
of $\lambda_F$ as they propagate for a certain length. In
our case this characteristic length is the length covered by quasiparticle
diffusing across the sample (see, e.g., \cite{Dirac}).

The procedure for doing the semiclassical mapping can be reduced to the
following.  
For a given point ${\bf r}$ in the N region we have to introduce all 
different trajectories which connect this point with   points on the 
barrier  (where the momentum of the quasiparticle is $\bf p$), and 
experiencing all possible sequences of scatterings induced by  N/S 
boundaries and barriers.
According to Ref.~\cite{Dirac}, the semiclassical wavefunction can be
constructed  as a sum of partial contributions $\Psi_n$ corresponding to
these trajectories, expressed in terms of the classical action 
$S_l=\int_{L_l} {\bf p}d{\bf l}$. As a result one has 
\begin{eqnarray}
u ({\bf r})&=&\sum_l A^{(e)}_l({\bf r})e^{iS_l^{(e)}/\hbar}
\nonumber \\
v ({\bf r})&=&\sum_l A^{(h)}_l({\bf r})e^{iS_l^{(h)}/\hbar} \, .
\label{wf} 
\end{eqnarray}
If a trajectory is split by interacting with a tunnel barrier or
N/S interface, the wavefunction along that trajectory
undergoes a transformation described by the scattering matrix
already mentioned.
The smoothly varying function $ A^{(h)}_l({\bf r})$  can be found from  
the continuity equation for the current density, which together with the 
Hamilton-Jacobi equation for the action, guarantees that
Eq.~(\ref{wf}) is a solution of the Schr\"{o}dinger
equation \cite{Dirac}. 
Furthermore, one can readily verify that the wavefunction (\ref{wf}),
constructed as a sum  over trajectories, satisfies the boundary conditions 
corresponding to the scattering at the barriers and interfaces.

\section{Electron-hole transmission at low energies}
 The wave function formally constructed in Eq. (\ref{wf}) does not permit us to
carry out concrete calculations in a general case. The reason lies in the
complications that arise due the bifurcations of trajectories as they
undergo  Andreev and normal reflections at the N/S boundaries. The situation is
drastically  simplified in the region of small energies, $E\ll E_{Th}$, where
reflections in the Andreev channel send the quasiparticle back along 
the incident trajectory. This implies that the trajectory bifurcations
disappear and the problem reduces to an analysis of one-dimensional 
quasiparticle motion along a single trajectory with centers for back (Andreev) 
and forward (normal) scattering.

A trajectory of this type is shown in Fig.~1. In this case the problem 
to find the wavefunction at the boundary 
[see Eq.~(\ref{reserv})] is reduced to a quantum scattering problem for 
the configuration shown in Fig.~1b. The points of reflection at the 
N/S boundaries and the points of scattering at the tunnel barrier are 
shown with black dots and  black bars, respectively. Propagation between these
points is coherent in both electron and hole 
channels, which is illustrated by dashed and solid lines of equal lengths. 
For quasiparticles with finite energy $E$ the 
phase gains along the electron and hole trajectories do not completely
cancel since the momenta are now different. The resulting decompensation 
effect is of order  $E/E_{Th}$.
\footnote{The semiclassical trajectories for an electron of energy $E$ 
and a reflected hole of energy $-E$ are to be considered identical 
since they separate by less than $\lambda_F$ while diffusing a distance $L_S$ 
if $E\ll E_{Th}$.}

Wave functions in the adjacent sections
of Fig.~1b are connected by a scattering
matrix describing Andreev  and normal reflections at the N-S boundary. The
problem of evaluating the electron-hole transmission can now be formulated as
follows.  An electron in the reservoir  arrives at the tunnel barrier  (solid
up-arrow in Fig.~1b), and we have to find the amplitude of the outgoing
wave function in the hole channel (dashed up-arrow). This
problem is reduced to solving a set of matching equations for
amplitudes  of electron $a_n^{(e)}$  and hole $a_n^{(e)}$ excitations in
every section of propagation between scattering points. A further
simplification follows as a consequence of the  resonant transmission caused by
multiple electron-hole transformations.  Such a resonance occurs, as we will
show, if the superconductor phase  difference $\phi$ is close to an odd multiple
of $\pi$ corresponding to a  large number of trajectories contributing to a
constructive interference between scattering events. If  $|r_N|\ll 1$,
where $r_N$ is the probability amplitude for non-Andreev (normal) 
reflection at an N/S boundary, the main
contribution comes from trajectories wich do not include successive reflections
at the same N/S boundary. Taking these observations into account the
following set of algebraic matching equations emerge,  
\begin{eqnarray}
\nonumber
a_0^{^{(e)}}&=&r_N^{(2)}e^{i\Phi^{(e)}_1}a_1^{(e)}+(1-\frac 12\epsilon
_r)r_A^{(2)}e^{-i\Phi^{(h)}_0}a_0^{^{(h)}} \\ \nonumber
a_1^{^{(h)}}&=&-r_A^{(2)*}e^{i\Phi^{(e)}_1 }
a_0^{^{(e)}}+r_N^{(2)*}e^{-i\Phi^{(h)}_0}a_0^{^{(h)}}
\\ \nonumber
a_{-1}^{^{(e)}}&=&r_N^{(1)}e^{i\Phi^{(e)}_0}a_0^{^{(e)}}+
r_A^{(1)}e^{-i\Phi^{(h)}_{-1}}a_{-1}^{^{(h)}}
\\ \label{setA}
a_0^{^{(h)}}&=&-(1-\frac 12\epsilon
_r)r_A^{(1)*}e^{i\Phi^{(e)}_0}a_0^{^{(e)}}+r_N^{(1)*}
e^{-i\Phi^{(h)}_{-1}}a_{-1}^{^{(h)}}-%
\sqrt{\epsilon _r}e^{i\Phi^{(e,1)}_0} \\ \nonumber
a_1^{^{(e)}}&=&r_N^{(1)}e^{i\Phi^{(e)}_2  }a_2^{^{(e)}}+
r_A^{(1)}e^{-i\Phi^{(h)}_1 }a_1^{^{(h)}}
\\ \nonumber
a_2^{^{(h)}}&=&-r_A^{(1)*}e^{i\Phi^{(e)}_2 }a_2^{^{(e)}}+
r_N^{(1)*}e^{-i\Phi^{(h)}_3}a_3^{^{(h)}}
\\ \nonumber
\ldots &=& \ldots
\end{eqnarray}

The amount of phase gained after propagation along the  trajectories in section
$n$ is 
$$
\Phi^{(e,h)}_n(E)  = \int_{L_n}p^{(e,h)}(E)dl/\hbar \approx \Phi_n\pm
\tau_n E/\hbar \, ,
$$ 
where $\Phi_n  =\Phi^{(e,h)}_n(0)$ and $\tau_n$ is propagation time
in section n. The quantities $r_A^{(1,2)}$ and $r_N^{(1,2)}$ are, respectively,
the probability  amplitudes for Andreev and normal reflection at N/S boundaries
1 and 2. Phases and amplitudes of an electron or hole along the semiclassical
paths are defined in such a way that no phase has been gained at the
beginning of a particular electron or hole section $n$. Hence the amplitude is 
 $a_n^{(e,h)}$ at the beginning of the section, and 
$a_n^{(e,h)}e^{\pm i \Phi^{(e,h)}_n}$ at its end.
The phase gain  between the tunnel barrier and the left N/S
boundary (N/S boundary 1) is denoted by $\Phi_0^{(e,1)}$. In Eq.(\ref{setA}) a 
coefficient $\epsilon _r\ll 1$ characterizing the coupling through
the tunnel barrier  has furthermore been introduced. We note that one can
show that the large phases $\Phi_n$ can be removed from the set of equations
(\ref{setA}). This is a manifestation of the fact that the electron and
hole phase gains compensate each other at $E=0$. 

According to our construction the probability 
$|C_{{\bf p}}({\bf r}_{\parallel})|^2$
of an electron-hole 
transmission at the point ${\bf r}_{\parallel}$ of the barrier 
 is related to the 
amplitude $a_0^{(h)}$  on the trajectory of Fig.~1 corresponding 
to injection at point ${\bf r}_{\parallel}$ as,
\begin{equation}
|C_{{\bf p}}({\bf r}_{\parallel})|^2=\epsilon_r|a_0^{(h)}|^2 \, .
\label{connetion}
\end{equation}
Equations (\ref{setA}) and (\ref{connetion}) together with (\ref{intcon}) give
the complete solution for the oscillatory, $\phi$-dependent part of the excess
conductance. 

When $\epsilon_r=0$, Eq.~(\ref{setA}) gives a set of Andreev levels 
$E_\alpha $. Therefore, one can expect that if $\epsilon_r \ll 1$ the 
trasmission probabilty $|C_{{\bf p}}^{(h)}({\bf r_{\parallel}})|^2$ is 
of the Breit-Wigner form. Indeed, as shown in Appedix 1, the transmission
probability $|C_{{\bf p}}^{(h)}({\bf r_{\parallel}})|^2$ in this limit
can be expressed as  %
\begin{equation}
|C_{{\bf p}^{(h)}({\bf r_{\parallel}})}|^2=
\frac{4\epsilon _r^2E_0^2|a_{0,\alpha}^{(h)}|^2}{4(E-E_\alpha )^2
+\epsilon _r^2 E_0^2 } \, .
\label{randprob1}
\end{equation}
Here $E_0$ is the spacing of Andreev levels generated by the trajectory 
when  $\epsilon_r=0$ and $n=0$,  $a_{0,\alpha}^{(h)}$ is the solution of 
Eq.~(\ref{setA}) for $E=E_{\alpha}$ and $\epsilon_r=0$.  Because of the random
variation of propagation times $\tau_n$, the
functions $a_{0,\alpha}^{(h)}$ are localized along the one-dimensional ladder
shown in Fig.~1b.

Further simplifications arise as one integrates the electron-hole 
transmission probability $|C_{{\bf p}}^{(h)}({\bf r_{\parallel}})|^2$ [see 
Eq.~(\ref{intcon})] over the area of the injector area and over the 
the Fermi surface in momentum space. This integration corresponds to 
summing over different  trajectories
and one can think of it as averaging over various distributions of 
$\tau_n$. It follows that $T(E)$ can be expressed as 
\begin{equation}
T(E)=N_{\perp}\left\langle\left\langle\frac{4\epsilon_r^2E_0^2|a_{0,\alpha}^{(h)}|}
{4(E-E_\alpha )^2+\epsilon _r^2 E_0^2 }\right\rangle\right\rangle \, ,
\label{avT}
\end{equation}
where $N_{\perp}= Sp_F^2/h$, $S$ is the area of the injector and
$\langle\langle\ldots\rangle\rangle$ 
implies an averaging over $\tau_n$. If $k_BT \gg \epsilon_r E_0$
  one can neglect the width of the 
resonance  for relevant energies, $E\sim kT$, and conclude that
\begin{equation}
T(E)=N_{\perp}\left\langle\left\langle\sum_{\alpha}\epsilon_r
|a_{0,\alpha}^{(h)}|^2 \delta([E-E_\alpha]/E_0)\right\rangle\right\rangle \, .
\label{avT1}
\end{equation}

The distribution of propagation times $\tau_n$ depends on the details of the
disordered potential in the mesoscopic normal region. These are not known, but
it is natural to assume that propagation times along different sections of the
semiclassical trajectory (see Fig. 1a) are uncorrelated. Under this assumption
one can, as detailed in Appendix 1, directly  express the transmission
probability $T(E)$ in terms of the average density of Andreev states coupled
with the  reservoir,\footnote{It is necessary to use the
fact that $\left\langle\left\langle|a_{n,\alpha}^{(h)}|^2
\delta(E-E_{\alpha})\right\rangle\right\rangle$ does not
depend on $n$ if the $\tau_n$'s are uncorrelated}
\begin{equation}
T(E)=N_{\perp}E_{Th}\left\langle\left\langle\nu(E)\right\rangle\right\rangle
\label{density}
\end{equation}
with
\begin{equation}
E_{Th} = \left\langle\left\langle E_0\right\rangle\right\rangle = 
h\left\langle\left\langle\tau^{-1}\right\rangle\right\rangle \, .
\label{Thouless}
\end{equation}
Here $\nu(E)$ is the density of Andreev states generated by a given 
semiclassical trajectory when $\epsilon_r=0$. 

In order to proceed with an analytical approach we choose a Lorentz form
for the distribution function $P(\tau)$, 
\begin{equation}
P(\tau)=\frac{1}{\pi}\frac \gamma {(\tau-\bar \tau)^2+\gamma ^2},  
\label{Lorentz}
\end{equation}
As shown in Appendix 2, this choice permits us to derive an  
analytical expression for the averaged density of state  
%
%
%
\begin{equation}
\left\langle\left\langle\nu (E)\right\rangle\right\rangle =
\frac{2|E|}\pi \int\limits_{-\infty}^\infty \frac{E_1^2\nu
_0(E_1)}{E_1^4+4E^4}dE_1 
\label{nufinal} 
\end{equation}
(we have used $\gamma =\overline{\tau}$, see \cite{GOdif}). The quantity  
$\nu _0(E)$ in the integrand of eq.~(\ref{nufinal}) is the density of 
states
of the periodic chain of Fig.~1b, if $\tau_n=\overline{\tau}$ for all $n$. 
It is straightforward to find this density of states to be
\begin{equation}
\nu _0(E)=\frac 2\pi \frac{|E|}{\sqrt{(E^2-E_{\min }^2)(E_{\max }^2-E^2)}} \, ,
\label{nuperiod}
\end{equation}
The energies $E_{\min }$ and $E_{\max }$ are lower and upper edges of
the  energy
band of the periodic chain of Fig.~1b. They can be expressed 
in terms of the ``hopping integral" $r_N^{1,2}$ as
\begin{equation}
E_{\min ,\max }=E_{Th}\sqrt{\delta \phi ^2+\left( |r_N^{(1)}|\pm
|r_N^{(2)}|\right)^2 }\quad \delta \phi = {\rm Min} |\phi -\pi(2k+1)| \, .
\label{band}
\end{equation}
Finally, using Eqs. (\ref{cond}), (\ref{density} and
\ref{nufinal})  one finds that the
low-temperature conductance  can be expressed as
\begin{eqnarray}
G &=&N_{\perp } \frac{2e^2}{h}\frac{\epsilon _r\sqrt{2}}{\overline T}
\int_0^\infty \frac x{%
\cosh ^2(x/2\overline T)}\times   \label{condfin} \\
&&\ \left\{ \frac{\sqrt{(4x^4+E _{min}^4)(4x^4+E _{max}^4)}+
E_{min}^2E_{max}^2-4x^4}{(4x^4+E _{min}^4)(4x^4+E_{max}^4)}\right\}
^{1/2}dx \, , \nonumber
\end{eqnarray}
where $\overline{T}=T/T_{Th}$ with $T_{Th}=E_{Th}/k_B$ defined by
Eq.(\ref{Thouless}).

The excess conductance $G(\phi)$ 
is plotted as function of phase difference
in Fig.~2, while the maximum oscillation amplitude $G_{max}(T)$ is
plotted as a function of temperature in Fig. 3. 
We emphasize two distinguishing features: 1) the excess conductance has
sharp maxima at $\phi = \pi(2k+1), 
\;k=0,\pm1,...$; 2) The peak oscillation amplitude has a maximum
value for a temperature much below the Thouless temperature,
$T\ll T_{Th}$, which qualitatively distinguish 
these results from those obtained for a completely transparant boundary 
between the mesoscopic region and the reservoirs.
Without potential barriers between the reservoirs and the mesoscopic normal
region there are no Andreev 
states that can contribute to the inter-reservoir transport. With 
such barriers present Andreev levels are well defined and long-lived.
The peak of the conductance in Fig. 2 is due to resonant tunneling
through a macroscopic number of such Andreev levels. 
A small amount of non-Andreev (normal) quasiparticle reflection at the N/S 
boundaries is not, as can be seen in Fig. 3, 
detrimental to the resonant tunneling effect. Rather it results in an
energy shift of the 
position of the resonance (provided the probability for normal
reflections at the two N/S boundaries are 
different). 
As a result the position of the maximum of the peak amplitude 
is shifted to a finite temperature, 
$T\sim \left||r_N^{(1)}|-|r_N^{(2)}|\right|T_{Th}$.
\begin{figure}
\centerline{\psfig{figure=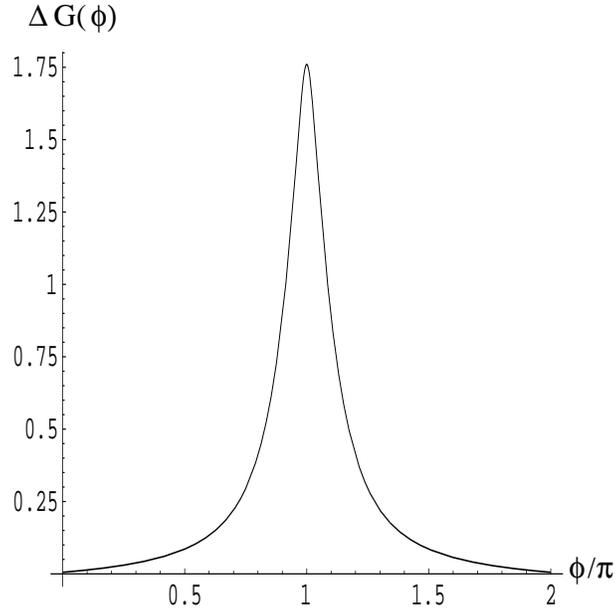,width=8 cm}}
\caption{ \label{fig:two}
Normalized excess conductance $\Delta G(\phi)/ G_N$ of the S/N/S structure
shown in Fig.~1a as a function of the phase difference $\phi$ between
the two superconductors. $\Delta G (\phi) \equiv G(\phi)
-G(\phi=0)$, $G_N =\epsilon_r (2e^2/h) N_{\perp}$; $T/T_{Th}=0.1$,
$r_N^{(1)}=0.1$, $r_N^{(2)}=0.2$ }
\end{figure}

\begin{figure}
\centerline{\psfig{figure=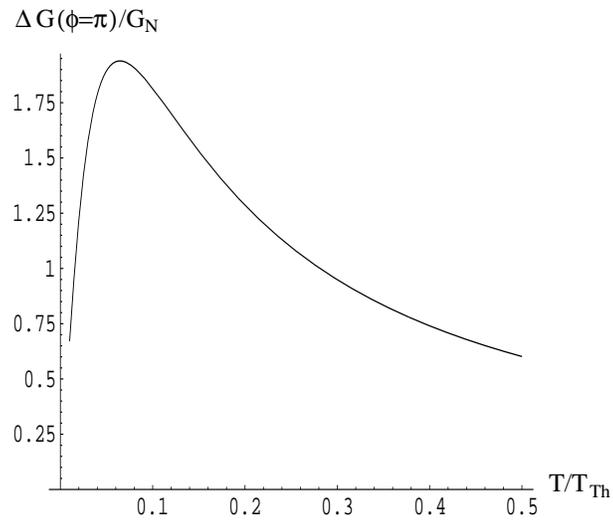,width=8 cm}}
\caption{ \label{fig:three}
Temperature dependence of the maximum normalized excess conductance 
$\Delta G(\phi=\pi)/G_N$ for the same tunnel barrier parameters as in Fig.~2. 
}
\end{figure}

\section{Conclusions}
We have shown that resonant tunneling through Andreev levels may give rise
to an excess quasiparticle contribution to the normal conductance of an
S/N/S sample at energies much below the Thouless energy and with maximal
amplitude when the phase difference between the superconductors is on
{\it odd} multiple of $\pi$.
That resonant tunnelling through Andreev levels could give rise to a
``giant'' effect was proposed for the case of a ballistic normal region
in Ref.~\cite{GO}.
The effect discussed here
is not sensitive to the details of the 
semiclassical motion of quasiparticles inside the disordered normal region.
A more remarkable fact is that the main conclusion about the importance of
resonant transmission through a macroscopically large number of Andreev 
levels  seems to be valid not only for a smooth 
disorder potential allowing a semiclassical analysis to be carried out as 
done here. Indeed, one can consider the density of states in an S/N/S 
junction isolated from reservoirs and solve the Eilenberger-Usadel equation, 
%
%
which is valid for a short-range disorder potential in the normal region.
This approach gives, according to Ref.~\cite{Spivak1}, a 
gap in the spectrum which closes at $\phi = \pi$ with the density of states   
diverging at $E=0$. 

In recent experiments \cite{Hideaki97,Rais} oscillations
of the conductance $G(\phi)$ have been observed, which are well described 
in the framework of the thermal effect of Nazarov and Stoof \cite{NazarovStoof}
at temperatures around the Thouless 
temperature, $T\sim T_{Th}$, but are significantly diferent from what the 
thermal effect can explain at lower temperatures.
The conductance maxima were found to be $\pi$-shifted in both 
experiments in the range of low temperatures. The maximum oscillation
amplitude $\Delta G (T)$ was observed at $T\approx 20$ mK (the Thouless 
temperature was $\approx 200$ mK \cite{Rais}). One may speculate (see 
\cite{Rais}) that grain boundaries and geometrical feauters of the 
contacts act to split the semiclassical quasiparticle 
trajectories in the same sense as the potential barriers in the model used 
here. If so one would expect the thermal effect 
\cite{NazarovStoof,VolkovLambert} and the resonant tunneling effect
to co-exist. In this 
case the cross-over in the phase and temperature dependences of the 
conductance oscillations could be understood as a result of competition 
between the``high temperature" thermal effect and the ``low temperature" 
resonant tunneling through Andreev levels. 

\vspace{2mm}
{\bf Acknowledgment.}
Support from the Swedish KVA and NFR and from the National Science 
Foundation under Grant No. 
PHY94-07194 is gratefully acknowledged. MJ is grateful for
the hospitality of the Institute for Theoretical Physics, UC Santa Barbara,
where part of this work was done.

\section*{Appendix 1}
\noindent
The set of equations (\ref{setA}) can be re-arranged in the
following way,
\begin{equation}
\left\{ \widehat{\omega }(\varepsilon )+\epsilon _r\widehat{P}\right\} 
|a\rangle = \sqrt{\epsilon _r}|g^{(e)}\rangle  \, .
\label{setM}
\end{equation}
Here $\widehat{\omega }(\varepsilon )=\widehat{1}-\widehat{U}(\varepsilon)$,
 $\widehat{1}$ is a unit matrix of order $2N\times 2N$ ($N$ is the
number of sections in the chain in Fig.~1b), and $\widehat{U}(\varepsilon )$ 
is a
$2N\times 2N$ unitary matrix whose explicit form can be
found from Eq.~(\ref{setA}) by setting $\epsilon _r=0$. The components
of the vector $|a\rangle$ are the amplitudes $a_n^{(i)}$, where the subscript 
$n$ labels the sections of the chain ($n=0,\pm 1,\pm 2,...\pm N/2$); the 
superscript
$i$ denotes the electron
($i=e$) and hole ($i=h$) paths,
 $\widehat{P}$ is an operator projecting onto
the section of injection  
\begin{equation} 
\widehat{P}=(1/2)\left\{|g^{(e)}\rangle\langle g^{(h)}|-|g^{(h)}\rangle\langle
g^{(e)}|\right\} \, .
\label{projection}
\end{equation}
The vector $|g^{(e)}\rangle$ has only one non-zero component $\delta
_{n,0}\delta _{i,e}$ as has  $|g^{(h)}\rangle$, $\delta _{n,0}\delta _{i,h}$
[see EQ.~(\ref{setA})].
The energy spectrum of a quasiparticle moving along the path of Fig.~1b in the
absence of any coupling to the reservoirs is determined by the roots of the
determinant of the matrix $\widehat{\omega }$.
In order to find the resonant transmission amplitude we apply resonant
perturbation theory to the set of
algebraic equations (\ref{setM}) assuming the barrier transparency 
$\epsilon _r$
to be small ($\epsilon _r\ll 1$) and the energy $E$ of the incoming 
qusiparticle
to be close to the energy level $E_\alpha $. Expanding the matrix 
$\widehat{\omega }(E)$ to lowest order in energy, $\widehat{\omega
}(E)=\widehat{\omega }(E_\alpha )+\widehat{\omega }^{^{\prime }}(E_\alpha
)\left(E -E_\alpha \right)$ and expanding the vector $|a\rangle$ 
to lowest order
in the small parameter $\sqrt{\epsilon _r}$ as 
\begin{equation}
|a\rangle=\gamma |e_\alpha\rangle+\sqrt{\epsilon _r}|a_\alpha^{(1)}\rangle,
\label{appexpand}
\end{equation}
one gets a set of algebraic equations,
\begin{equation}
\widehat{\omega }(E_\alpha )|a_\alpha ^{(1)}\rangle=-\gamma \left\{ \left(
E -E_\alpha \right) \widehat{\omega }^{^{\prime }}(E_\alpha
)+\epsilon _r\widehat{P}\right\} |e_\alpha\rangle+\sqrt{\epsilon _r}|g\rangle
\label{firstappr}
\end{equation}
where the vector $|e_\alpha\rangle$ is a normalized non-trivial solution of the 
equation\footnote{Here and below, there is no summation with respect to double
indices.} 
\begin{equation}
\widehat{\omega }(E_\alpha )|e_\alpha >=0.  
\label{zeroappr}
\end{equation}
The constant $\gamma $ is determined by the condition that Eq.~(\ref
{firstappr}) has nontrivial solutions. Its value can be found by multiplying 
both sides of the set of equations with $\langle e_\alpha |$, the latter vector
being a non-trivial solution of the set of equation,
\begin{equation}
\langle e_\alpha|\,\widehat{\omega }(E_\alpha )=0  \label{zeroreversed}
\end{equation}
(the vectors $\langle e_\alpha|$ and $|e_\alpha \rangle$ are normalized in such a
way that $\langle e_\alpha |e_\alpha \rangle=1$). As a result, one gets
\begin{equation}
\gamma =\frac{\sqrt{\epsilon _r}\langle e_\alpha |g^{(e)}\rangle}
{\langle e_\alpha|\,\widehat{\omega}^{^{\prime }}(E_\alpha )|e_\alpha
\rangle\left( E -E_\alpha \right) +\epsilon _r\langle e_\alpha
|\widehat{P}|e_\alpha \rangle}  \, ,
 \label{gamma} 
\end{equation}
with $\langle e_\alpha |g^{(e)}\rangle = a^e_{0,\alpha}$.
As we are considering the case of low normal reflection probability amplitudes, 
$|r_N^{(i)}|\ll 1$, and the terms in the denominator already contain small
factors $\left( E -E_\alpha \right)$, 
the amplitudes $r_N^{(i)}$ can be neglected in these
terms to lowest order. This gives as a result that $\langle e_\alpha
|\widehat{\omega }^{^{\prime }}(E_\alpha )|e_\alpha \rangle =i/E_0$ ($E_0=\hbar
v_F/\tau_0$, $\tau_0$ is the  propagation time along the path in the
section of injection in Fig.~1b) and  $\langle e_\alpha |\widehat{P}
|e_\alpha \rangle=1/2$.

One can show that in the
resonant approximation the second term on the right hand side of
Eq.~(\ref{appexpand}) can be neglected  if the prameter $\epsilon _r\ll 1$.
Hence the amplitude of the electron-hole transition is $a_0^{(h)}=\gamma
\langle g^{(h)}|a\rangle$. Therefore, using  Eq.~(\ref {connetion}) and
(\ref{gamma}) we get the Breit-Wigner formula  (\ref{randprob1}).

\section{Appendix 2}
As shown by Slutskin \cite{Slutskin}, the density of states of a 
one-dimensional chain of the type in Fig.~1b can be written as a Fourier 
series,
\begin{equation}
\nu (E)=\frac{1}{hN}\sum\limits_{n=1}^N\tau_n\left( F_{n}+F_{n}^{*}+1\right)
\, ,
\label{Fourier}
\end{equation}
where
$$
F_{n}=\sum\limits_{\{ l_m \}}D(\{ l_n \})\prod_{m}
e^{i l_n \tau_m E/\hbar} \, .
$$
Here $\{ l_n\}$ labels various sets of numbers $(l_{1,}l_{2,...}l_N)$ 
where $l_{n=1}=1$ 
 and $l_{m\not= n}$ is equal to 0 or 1;
$D(\{ l_n \})$
are Fourier coefficients which depend on the ``hopping integrals" 
$r_N^{(1,2)}$.
Since the ``Fourier amplitudes" do not depend on $\tau_n$
one only has to average the product  
$\tau_n\prod_{n}{\rm e}^{il_n \tau_n E/\hbar}$ while 
calculating the average density of
states. Using the Lorentzian distribution (\ref{Lorentz}) one finds the result
\begin{equation} 
\left\langle\left\langle \tau_n\prod_{m}
{\rm e}^{i l_n \tau_m E/\hbar}  \right\rangle\right\rangle
=
\frac{\gamma}{\pi  \overline{\tau}} \int\limits_{-\infty }^\infty 
dE_1\frac{E_1
}{(E_1 -E)^2 + (\gamma/\overline{\tau})^2E^2}\left(\overline{\tau}\prod_{n}
e^{i l_n \overline{\tau} E_1/\hbar}\right)
  \label{nufinal1}
\end{equation}
The term in parenthesis in this expression
is exactly the same  as what appear on the right hand side of
Eq.~(\ref{Fourier})
provided the system in Fig.1b is periodic with all $\tau_n=
\overline{\tau}$. 
 From this and the fact that $\nu_0(E_1) = \nu_0(-E_1)$ 
the result Eq.~(\ref{nufinal}) follows immediately.

\end{document}